\documentstyle[graphicx,amsfonts,amssymb,11pt]{article}

\newcommand{\dfr}[2]{\frac {\displaystyle #1}{\displaystyle #2}}

\begin{document}
\title{Structural transitions and non-monotonic
relaxation processes in liquid metals}
\author{M.~G.~Vasin\footnote{%Corresponding author
vasin@udm.net}, V.~I.~Lad'yanov}
%\address{ Physical-technical Institut, Ural Branch of Russian
%Academy of Sciencies, 132 Kirov st., 426000  Izhevsk, Russia\\
%e-mail: vasin@udm.net}

\maketitle
\begin{abstract}
Structural transitions in melts as well as their dynamics are
considered. It is supposed that liquid represents the solution of
relatively stable solid-like locally favored structures (LFS) in
the surrounding of disordered normal-liquid structures. Within the
framework of this approach the step changes of liquid Co viscosity
are considered as liquid-liquid transitions. It is supposed that
this sort of transitions represents the cooperative medium-range
bond ordering, and corresponds to the transition of the "Newtonian
fluid" to the "structured fluid". It is shown that relaxation
processes with oscillating-like time behavior ($\omega \sim
10^{-2}$~$s^{-1}$) of viscosity are possibly close to this point.
\end{abstract}
%\draft \pacs{61.20.Lc, 61.20.Cy, 64.70.Ja}

\section{INTRODUCTION}
It is known, that properties of both crystalline and amorphous
substances often depend on their production conditions, and in
many respects the initial liquid state determines these
properties. At the same time, the liquid state of many substances
is insufficiently investigated now. In recent years it has been
found, that unusual properties and phenomena are characteristic
for many fluids, but it is difficult to explain these properties
within the framework of the elementary representations of the
structureless liquids. One of the most interesting challenging
phenomena being studied is the "liquid--liquid phase transition".
At present the greatest progress is achieved in examination of
network-forming liquids (such as water). Besides the papers, which
inform about observation of similar phenomena in simple
single-component systems, such as P~\cite{Ca} or C~\cite{Cb}, have
appeared. A possibility of existence of liquid--liquid transitions
in the metal melts is also being discussed~\cite{Dah, Cd}.

Whereas at present the existence of liquid--liquid transitions in
the glass-forming melts is beyond question, their existence in
ordinary single-component metal liquids seems to be surprising. In
many respects it is because of lack of direct structure
observations. Nevertheless, there are already indirect evidences
that point to the existence of appreciable changes in the melts
structures. In particular, it is the appearance of "structural
fluid"-like behaviour of some single-component melts at
temperature decreasing, that appears as a little step change of
liquid viscosity which goes on with the change of the viscosity
activation energy~\cite{Dah, v2}.

Besides, there are facts, which suggest that the process of
establishing a thermodynamic equilibrium in the system close to
this point can have a non-monotonic oscillating
character~\cite{B2}. This effect was observed on the polytherms of
viscosity, surface tension and magnetic susceptibility of metal
melts, and the low-frequency modes $\omega \sim 10^{-2}$ $s^{-1}$
predominated in these oscillations~\cite{B2, Os} several times.
For example, the time dependence of the kinetic viscosity of
liquid Co at $T=1600$ $^{\circ }$C is presented in Fig.1(b). These
results were obtained in~\cite{v1} by the torsional vibrations
method in isothermal regime. As usual~\cite{v1}, the temperature
was rapidly elevated and maintained at the given level. In the
figure one can see considerable fluctuating of the measured
quantities of viscosity. They have a stochastic character, and
their amplitude exceeds the experimental error. It is important
that the system temperature is close to the point of the step
change of viscosity (Fig.2). The dispersion of viscosity values in
this point considerably exceeds the dispersions of viscosity
values at other temperatures. The usual variation of viscosity is
$\Delta \nu \sim 0.7\cdot 10^{-7}$ m$^2/$s for discussed
experiments, but this dispersion grows by 3--4 times near the
vicinity of the temperature at which the viscosity polytherms jump
are observed. The spectral-correlation analysis of these data was
shown in the following papers~\cite{Vasin1, Vasin2}. The results
have confirmed the supposition that a pronounced low-frequency
mode is present in these data~(Fig.3).

Unfortunately, so far, the reliability of the above data has been
questionable and true conclusions about the character and
mechanism of the observed relaxation processes have remained
unsolved. It is connected with the difficulties of developing
specific precise experimental techniques and processing the
obtained data as well as with the lack of the appropriate
theoretical models. However, it is natural to suppose that these
effects can be generated by some local structure modifications.
Since the oscillation relaxation processes are observed in the
neighborhood of the step change of liquid viscosity we believe
these effects to be connected with each other. Besides these
vibrations look like as a fluctuations reinforcement near the
phase transition point (Fig.2). Therefore in this paper we assume
that liquid--liquid transition point is root of both the step
change of liquid viscosity and the oscillating relaxation.

\section{THEORETICAL MODEL}
\label{TM} It is known that thermodynamic systems are metastable
at the temperatures close to critical ones. This metastability can
be described by addition of the non-linear part to non-equilibrium
thermodynamic potentials of the system. On the other hand, such
systems undergo strong thermodynamic fluctuations as well as the
effect of external ''noise'' because of both the influence of the
experimental plant and thermal fluctuations. Thus, one can suppose
that the combination of strong none-linearity of the metastable
system and fluctuations is the reason of the oscillating
relaxation, but the nature of metastability is clear only in the
case of melts, whereas the nature of the liquid--liquid transition
in liquid metals is still unclear. Therefore, we think, that the
problem of theoretical description of the oscillating relaxation
is reduced to two problems: first of all, it is necessary to
define the nature of the liquid--liquid transition in the metal
liquid and give its theoretical model. After that one can discuss
the dynamic peculiarity of this model, and the chance of
appearance of the low frequency mode in the oscillation spectrum.

In most of theoretical papers liquid is treated as the one
consisting of adjoining to each other and interacting with each
other elementary local volumes which include atoms of several
coordination spheres~\cite{Tanaka, Pat, Son1,  Bakai}. It is
expected that each of these local volumes has only a few
energy--profitable configurations of the nearest order (local
states). The long-range order is absent here as the local groups
of atoms can be oriented differently with respect to each other.
The topological structures of these locally--ordered formations
have some symmetry. One can suppose that this local symmetry
corresponds either to the crystal symmetry (solid-like), e.g.
f.c.c. or b.c.c., or to the icosahedral symmetry (normal-liquid).
We believe that as the temperature decreases both normal-liquid
and solid-like locally favored structures (LFS)~\cite{Tanaka} can
form clusters. These clusters have a finite size and continuously
transform, since the thermal fluctuations of the system lead to
continuous destroying old bonds and to simultaneous forming new
ones.

Let us consider the case when the temperature of the melt is close
to that of the liquid--liquid transition. The foolproof theory of
the structural transitions in liquids is not available for the
time being. Therefore to describe the metastable properties of the
close to this transition system we will consider the simplest
model. As well as it was in~\cite{Tanaka} we will describe the
close to liquid-liquid transition system as the system that is
near the gas--liquid transition. We believe that the order
parameter is connected with density. Therefore, using the density
as the order parameter we introduce the following free energy,
which governs $\psi $ fluctuations near the gas--liquid-like
critical point $T_{cr}$~\cite{Klim}:
\begin{equation}\label{01}
  F_0=\displaystyle \frac 12\int d^dx\left[ \;(\vec \nabla \psi )^2+\gamma (T,\,P) \psi +\alpha
(T-T_{cr}) \psi ^2
  +\frac {g}{2}\psi ^4\;\right] ,
\end{equation}
$\gamma (T,\,P)$ is the function of temperature $T$ and pressure
$P$, the parameters $g$ and $\alpha $ are weakly depending on
temperature.

\section{DESCRIPTION OF THE RELAXATION}
\label{DR} Let us consider the fluctuations of the order parameter
of metastable liquid in the region of phases co-existence ($\gamma
\approx 0$). In order to describe the relaxation dynamics of this
system we will consider the H-model~\cite{Halperin}. This model is
represented by the following equations system:
\begin{equation}\label{f5}
 \begin{array}{c}
   \displaystyle \dfr{\partial \psi}{\partial t}= \displaystyle \lambda \nabla
^2\dfr{\delta F}{\delta \psi }-
   g_0\vec \nabla \psi
   \cdot \dfr{\delta F}{\delta \vec v}+\theta ,\\[12pt]
   \displaystyle \dfr{\partial \vec v}{\partial t}=
   \displaystyle P^{\perp}\left[ \; \eta _0 \nabla ^2 \dfr{\delta F}{\delta \vec
v}+
   g_0\vec \nabla \psi \dfr{\delta F}{\delta \psi }
   +\vec \xi \;\right] ,\\[12pt]
   F=
   \displaystyle F_0-\int d^d x \left[ \;h(\vec x,\,t)\psi +\vec A(\bar
x,\,t)\cdot \vec v\;\right]
   ,\\[10pt]
   F_0=\displaystyle \frac 12\int d^dx\left[ \;\tau \psi ^2+(\vec \nabla \psi
)^2+\frac {g}{2}\psi ^4+\vec
   v^2\;\right],
 \end{array}
\end{equation}
where $\tau =\alpha (T-T_{cr})$, $h$ and $\vec A$ are
infinitesimal applied fields. The first equation describes the
dynamics of the order parameter $\psi $, and the second one
describes the dynamics of the transverse part of the momentum
density $v$, ($P^{\perp} $ is a projection operator which selects
the transverse part of the vector in brackets, $\lambda $ is the
coefficient of self-diffusion, $\eta _0$ is the viscosity,
$g_0=\left. 1\right/ \lambda \eta _0$ is the mode-coupling vertex,
the functions $\vec \xi $ and $\theta $ are the Gaussian white
noise source:
\begin{eqnarray*}
  \begin{array}{c}
   \langle \theta(x,\,t)\theta(x',\,t')\rangle=-\lambda \nabla
   ^2\delta(x-x')\delta(t-t'),\\[12pt]
   \langle \xi _i(x,\,t)\xi _j(x',\,t')\rangle=-\eta _0 \nabla
   ^2\delta(x-x')\delta(t-t')\delta _{ij}.
  \end{array}
\end{eqnarray*}
The critical properties of this model are known~\cite{Halperin}.
In particular, it is well known that the viscosity of such a
system depends on the character time of the experiment $
t_{ch}=1/{\omega _{ch}} $. In the case of the results discussed
above, this quantity is the frequency of torsional vibrations.
Therefore, one can really expect that the change of the viscosity
around $T_{cr}$ will be prolonged in some temperature interval
$T_1>T>T^{*}(\omega )$ rather than jump-like ($T_1$ is the
temperature of formation of the metastable phase). It agrees with
the experimental observations. Note, that this interval depends on
the frequency of torsional vibrations $\omega $:
\begin{equation}\label{f6}
  \displaystyle T^{*}(\omega )-T_{cr}\sim \left(\dfr{\omega }{\lambda
}\right)^{\left. 1\right/ (\nu
  z)}
\end{equation}
($\nu $ and $z$ are corresponding static and dynamic critical
exponents). At the low-frequency $\omega < 1/t_{cl}$ ($t_{cl}$ is
the life-time of the clusters ) the value of viscosity $\eta
(\omega )$ will be slightly dependent on the correlation length
$\xi _{c}$ (clusters size). But at the higher frequency $\omega
\gg 1/t_{cl}\sim \xi _c^{-z}$ the response to the mechanical
perturbation will be determined by the scale which is smaller than
the cluster size, and it will not depend on the frequency.

\section{THEORETICAL EXPLANATION OF THE VISCOSITY TIME OSCILLATING}
\label{TEO} Usually model (\ref{f5}) is used for theoretical
description of critical dynamics of gas-liquid transitions and
transitions in binary fluids. The system inertia is not taken into
account. The point is, the critical dynamics is usually
investigated, and in this case the rescaling operator increases
the importance of $\omega $, relative to $\rho \omega ^2$, and
after the renormalization procedure the term $\rho \omega ^2$ may
be neglected. However, we will investigate the nonlinear
stochastic system which is not found correct at the critical
point. In this case the non-linearity of the system can lead to
the state when even its weak inertia will essentially influence
long-time dynamics~\cite{Koffi}. In order to take it into account,
it is necessary to add the proportional to double $t$-derivation
term to the first equation of the system:
\begin{eqnarray*}
 \displaystyle \rho \dfr{\partial ^2\psi }{\partial t^2}+\dfr{\partial
\psi}{\partial t}= \displaystyle
 \lambda \nabla ^2\dfr{\delta F}{\delta \psi }-g_0\vec \nabla \psi
 \cdot \dfr{\delta F}{\delta \vec v}+\theta .
\end{eqnarray*}
To analyze this stochastic model, one can employ the standard
method of the stochastic deriving functional \cite{VS} and theory
of perturbation. According to these methods, the correspondent
field model will be described by a set of basic $\{ \psi ,\, v \}
$ and supplementary $\{\psi ',\, v' \} $ fields, and the effective
action will have the form (Fig.4) of:
\begin{eqnarray*}
\begin{array}{rcl}
S (\Phi) &=& -\lambda \psi ' \partial ^2 \psi ' + \psi ' [-\rho
\partial _t ^2 \psi-
\partial _t \psi - \lambda \partial ^2 ( \partial ^2 \psi - \tau \psi - \gamma \psi ^2-g \psi
^3 )
-\\[12pt]
&-& v \partial \psi ]+ \lambda ^{-1} g _{0} ^{-1} v ' \partial ^2
v ' + v ' [-\partial _t v + \lambda ^{-1} g _{0} ^{-1} \partial ^2
v + \psi
\partial (\partial ^2
\psi)].
\end{array}
\end{eqnarray*}
The propagators of fields $\psi $ and $v$ have the form of:
\begin{eqnarray*}
\begin{array}{c}
  \displaystyle \langle \psi \psi ^{\prime }\rangle =\langle \psi ^{\prime
} \psi \rangle
  ^T=\displaystyle \dfr {1\left/\rho \right.}{-\omega ^2-ia\omega +\varepsilon
  _k},\quad
  \langle \psi \psi \rangle =\dfr {2\lambda k^2\left/ \rho ^2\right.}{ \left(
-\omega ^2-ia\omega +\varepsilon
  _k \right) \left( -\omega ^2+ia\omega +\varepsilon
  _k\right)},\\[12pt]
  \displaystyle \langle v v^{\prime }\rangle =\langle v ^{\prime
}v\rangle
  ^T=\displaystyle \dfr {\lambda g_{0}P^{\perp}}{-i\omega \lambda g_0
+k^2},\qquad
  \langle v v\rangle =\dfr {2\lambda g_{0}k^2}{ \left( -i\omega \lambda g_0
+k^2\right) ^2},
\end{array}
\end{eqnarray*}
where
\begin{eqnarray*}
  \displaystyle \varepsilon _k=\dfr{\lambda }{\rho }\,k^2(k^2+r_0 ), \qquad
  a=\dfr{1}{\rho }\, ,
\end{eqnarray*}
and $r_0 $ is a renormalized quantity $\tau $, $k^2$ is the
impulse $\vec k$ squared. Below we will consider only the
propagators of the conservative order parameter field $\psi $. For
calculation it is convenient to use the $(k,\,t)$-representation,
in this case, the propagators become:
\begin{eqnarray*}
  \begin{array}{rl}
  & \displaystyle \langle \psi \psi ^{\prime }\rangle = \dfr {2\pi \,e^{-\left.
at\right/2}}{\rho \sqrt{-4\varepsilon
  _k+a^2}}\exp \left( -\dfr 12\sqrt{-4\varepsilon
  _k+a^2}|t|\right) ,\\[12pt]
   & \displaystyle \langle \psi \psi \rangle = \dfr {\lambda \pi k^2}{a\rho
^2\varepsilon _k} \,\exp\left(-\dfr 12\sqrt{-4\varepsilon
   _k+a^2}|t|\right)\left[
   \dfr {e^{\left.a|t|\right/2}}{\sqrt{-4\varepsilon
_k+a^2}}\left(\sqrt{-4\varepsilon _k+a^2}+
   a\right)-\right.  \\[12pt]
   & \displaystyle
   - \left.\dfr {e^{-\left.a|t|\right/2}}{\sqrt{-4\varepsilon
_k+a^2}}\left(\sqrt{-4\varepsilon
   _k+a^2}-
   a\right)\right],
  \end{array}
\end{eqnarray*}
when~$4\varepsilon _k<a^2$, and
\begin{eqnarray*}
  \begin{array}{rl}
  & \displaystyle \langle \psi \psi ^{\prime }\rangle = \dfr {2\pi
  \,e^{-\left. at\right/2}}{\rho \sqrt{4\varepsilon
  _k-a^2}}\sin \left( \dfr 12\sqrt{4\varepsilon
  _k-a^2}\,\,t\right) \left[ \theta (t)-\theta (-t)\right] ,\\[14pt]
  & \displaystyle \langle \psi \psi \rangle = \dfr {\lambda \pi k^2}{a\rho
^2\varepsilon _k} \dfr {\left[\theta(t)-
    \theta(-t)\right]}{\sqrt{4\varepsilon _k-a^2}}\left[ e^{-\left.
at\right/2}\left\{ \sqrt{4\varepsilon
   _k-a^2}\cos \left(\dfr 12\sqrt{4\varepsilon _k-a^2}\,\,t\right)
   +\right. \right.   \\[14pt]
  & \displaystyle \left.+ a\sin \left(\dfr 12 \sqrt{4\varepsilon
_k-a^2}\,\,t\right) \right\}
    + \,e^{\left. at\right/2}\left\{ a\sin \left(\dfr 12 \sqrt{4\varepsilon
_k-a^2}\,\,t\right)- \right.
    \\[14pt]
  & \displaystyle \left. \left. -\sqrt{4\varepsilon _k-a^2}\cos \left(\dfr
12\sqrt{4\varepsilon _k-a^2}\,\,t\right)
   \right\} \right] ,
  \end{array}
\end{eqnarray*}
when $4\varepsilon _k >a^2$. It is important that
the effective viscosity depends on $\omega $, and it can be
represented as
\begin{eqnarray*}
  \eta (\tau,\,\omega )=\displaystyle \eta _0\left[1 -\dfr {\displaystyle \lambda g_0
  \Sigma _{v'v}}{\displaystyle p^2}\right] ,
\end{eqnarray*}
where $\Sigma _{v'v}$ is a coupled-mode contribution to the
response function, $\vec p$ is an external impulse. In an one-loop
approach it can be represented in the diagram form (Fig.5), in
mathematic $(k,\,t)$-representation this contribution has the form
of:
\begin{eqnarray*}
\begin{array}{rcl}
  \Sigma _{v'v}(t,\,p)&=&\displaystyle \dfr {\pi ^2}{a\rho
^2}\displaystyle \int \dfr {d^3k}{(2\pi
  )^3}\dfr {[k_iP_{ij}^{\perp }k_j][p^2-2\vec p\cdot \vec k]}{[q^2+r_0]
\sqrt{-4\varepsilon
  _k+a^2}}\exp \left( -|t|\left[ \sqrt{-4\varepsilon _q+a^2}+\sqrt{-4\varepsilon
  _k+a^2}\right]
  \right)\times \\[12pt]
  & \times & \displaystyle \left[ \dfr a{\sqrt{-4\varepsilon _q+a^2}}\left(
1+e^{-at}\right)
  +\left( 1-e^{-at}\right) \left( \theta(t)-\theta(-t)\right)
  \right], \qquad (\bar q=\bar p-\bar k),
  \end{array}
\end{eqnarray*}
in case of $4\varepsilon _k <a^2$, and
\begin{eqnarray*}
\begin{array}{rcl}
  \Sigma _{v'v}(t,\,p)&=&\displaystyle \dfr {\pi ^2}{a\rho
^2}\displaystyle \int \dfr {d^3k}{(2\pi
  )^3}\dfr {[k_iP_{ij}^{\perp }k_j][p^2-2\vec p\cdot \vec
k]}{[k^2+r_0]\sqrt{(4\varepsilon
  _q-a^2)(4\varepsilon _k-a^2)}}\left[\theta(t)-\theta(-t)\right]\sin \left(
\dfr
12\sqrt{4\varepsilon _q-a^2}\,t \right)\times \\[12pt]
  & \times & \displaystyle \left[ e^{-at}\left\{ \sqrt{4\varepsilon
   _k-a^2}\cos \left(\dfr 12\sqrt{4\varepsilon _k-a^2}\,\,t\right)
   + a\sin \left(\dfr 12 \sqrt{4\varepsilon _k-a^2}\,\,t\right)
   \right\}+ \right. \\[14pt]
  & &\displaystyle + \left. \left\{ a\sin \left(\dfr 12 \sqrt{4\varepsilon
_k-a^2}\,\,t\right)-\sqrt{4\varepsilon _k-a^2}
  \cos \left(\dfr 12\sqrt{4\varepsilon _k-a^2}\,\,t\right) \right\}
  \right], \\[12pt] &&(\bar q=\bar p-\bar k),
  \end{array}
\end{eqnarray*}
in case of $4\varepsilon _k >a^2$. If we switch
from integration at $k$ to integration at $\omega
=\sqrt{4\varepsilon _k-a^2}$ in limit $t\to \infty $ and $p\to 0$
we can obtain:
\begin{eqnarray*}
\displaystyle \lim _{p\to 0}\dfr {\Sigma _{v'v}(t,\,p)}{p^2}=\dfr
{\left(\rho \lambda \right)^{-3/4}}{16\sqrt{2}a}\int d\omega
\left[ f_1(\omega ) \sin(\omega t)+f_2(\omega ) (\cos(\omega t)+1)
\right] ,
\end{eqnarray*}
where
\begin{eqnarray*}
\begin{array}{c}
\displaystyle f_1(\omega )=\dfr {\left( -r_0\sqrt {\lambda \rho
}+\sqrt {\lambda \rho r_0^{2}+w^2 \rho^2+1}\, \right)
^{3/2}}{\left( r_0\sqrt { \lambda \rho }+\sqrt {\lambda \rho
r_0^{2}+w^2 \rho^2+1}\right) \sqrt
{\lambda \rho r_0^2+w^2 \rho^2+1}}\,,\qquad f_2(\omega )=\dfr {f_1(\omega )}{\rho \omega }\,.%\\[12pt]

\end{array}
\end{eqnarray*}
Fig.6 shows the qualitative aspect of this spectrum. One can see
tow peaks in this figure. The $\omega =0$ peak corresponds to the
Goldstone mode and it is explained by the presence of preservation
law for $\psi $ and $v$. Another peak
\begin{eqnarray*}
\omega \simeq \dfr 1{\rho }\sqrt{1-\dfr{68.4\lambda \rho }{\xi _c
^4}}
\end{eqnarray*}
corresponds to the low-frequency oscillations of the system. Thus,
it is possible to conclude that the observed oscillations are
noise-induced, analogous to the noise-induced oscillations in
bistable systems, and excitation of the low-frequency modes in the
system can be a noise-induced transition \cite{Horst}
(non-equilibrium phase transition).

Then the condition of observation of these oscillations is $ \xi
_c \approx \displaystyle \left[ 68.4 \lambda \rho \right] ^{\frac
14 }, $ and one can estimate the correlation length quantity:
Since $\lambda =D_a a_0 ^2$, where $D_a \sim 10^{-9}$ $\left.
m^2\right/ s$ is the diffusion coefficient, $a_0\sim 10^{-9}$ $m$
is the interatomic distance, and $\rho $ is the relaxation time,
we believe that these values are of the order of the torsional
vibration period (it is $\rho \sim 1$ $s$ in our experiments
\cite{v1}). Then the correlation length is $\xi _c \sim 10^{-7}$
$m$. One can anticipate that it is the size of the flickering
clusters of the metastable phase. It should be noted that this
value is close to the size of the "Fischer clusters" which was
observed in supercooled liquids~\cite{Dah}, but direct observation
of these formations at the relatively high temperatures is not
available for the time being.

\section{CONCLUSIONS}
It is well known that liquid is a nonuniform system, and its
structure is characterized by the presence of the structure
clusters in it. Recently the polymerization-like processes have
been observed in metal systems by small-angle neutrons dispersion
experiments~\cite{Dah}. In these experiments the snowflake-like
large-size heterogeneity was discovered. We believe that similar
processes of liquid--liquid transition lead to appearance of
structured fluid (rheology) properties in liquids at relatively
low temperatures, and to the jumps in the viscosity polytherms of
some liquid metals.

In an attempt to explain the observed oscillations of viscosity,
we assume that presence of liquid--liquid transition point is the
possible root of appear of the properties oscillating in the
explored systems. We believe that they are caused by the exterior
noise. Nonlinearity of the system in the region of its
metastability is the reason of the increase of the fluctuations
dispersion, and the possible reason of the oscillating character
of these fluctuations. The latter is caused by the low-frequency
peak in the spectrum, and we believe that this effect analogous to
well-known noise-induced transitions in bistable
systems~\cite{Horst}. We would like to mark that one ought not to
consider the period of the viscosity oscillation as a life time of
the metastable subsystems (clusters). The sizes of the such
subsystems are about correlation radius $\sim \xi _c$ and their
fluctuations frequency is a reciprocal value for the lifetime
$t_{cl} \sim \xi _c^{z}\sim 10^{-10}$ s. As indicated above the
viscosity oscillation is the dynamic effect, which is determined
by influence of the fluctuating moving to the hydrodynamic moving,
and inheres in the whole system.

Further examination of nontrivial dynamic properties of such
systems will allow us to confirm or discard our hypothesis.

This study was supported by the RFBR grant (01-02-96455
r2001ural).

%\begin{references}

\begin{figure}[h]
   \centering
   \includegraphics[scale=1]{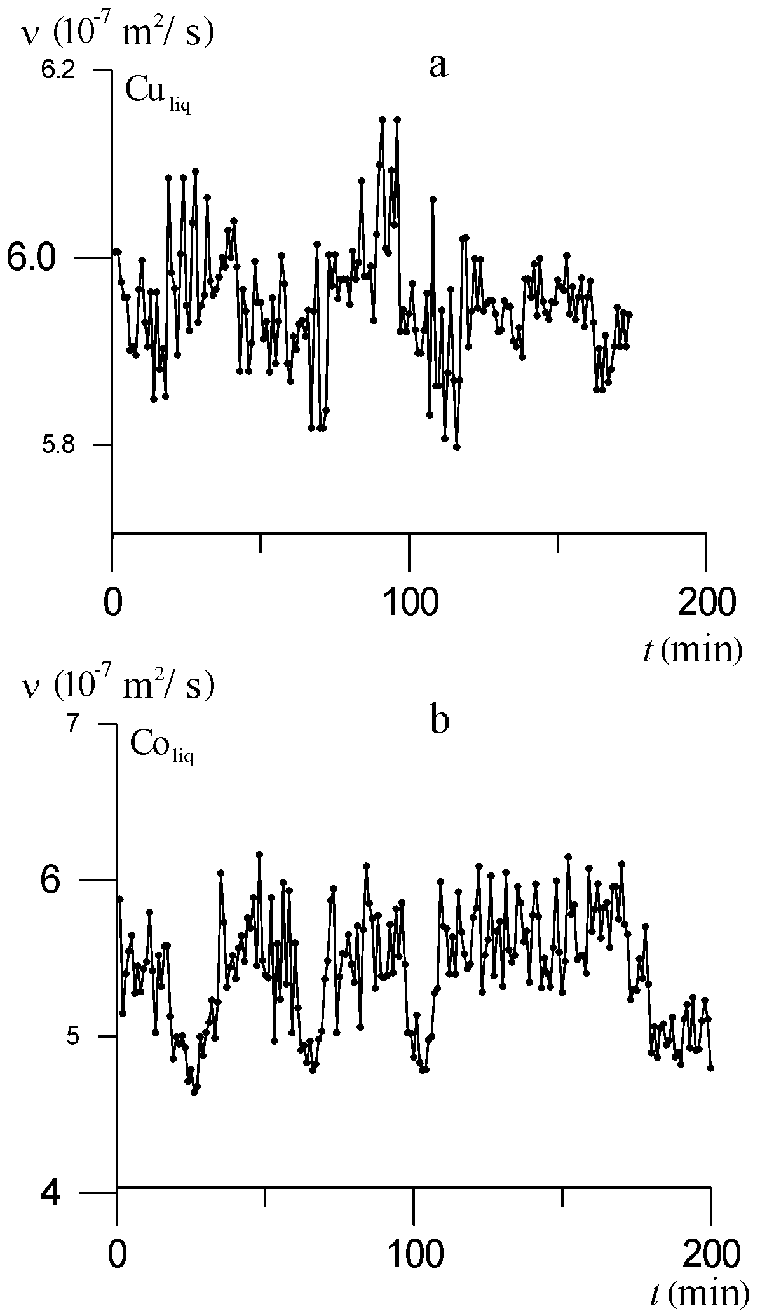}\caption{Time dependences of kinematic viscosity of liquid copper
\cite{Vasin1, Vasin2}(a) and liquid cobalt (1870 {\rm K}) \cite{v1} (b) close to the temperature of step change of viscosity.}
\end{figure}

\begin{figure}[h]
\centering
\includegraphics[scale=1]{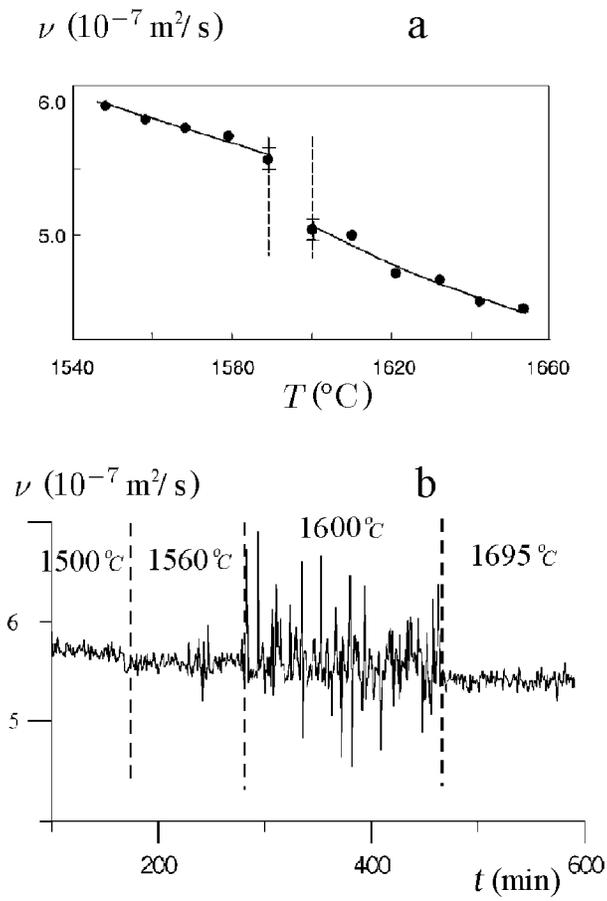}\caption{Temperature dependence of kinematic viscosity of liquid
cobalt (a) and time dependence's of liquid cobalt viscosity at various temperatures (b).}
\end{figure}

\begin{figure}[h]
   \centering
   \includegraphics[scale=1]{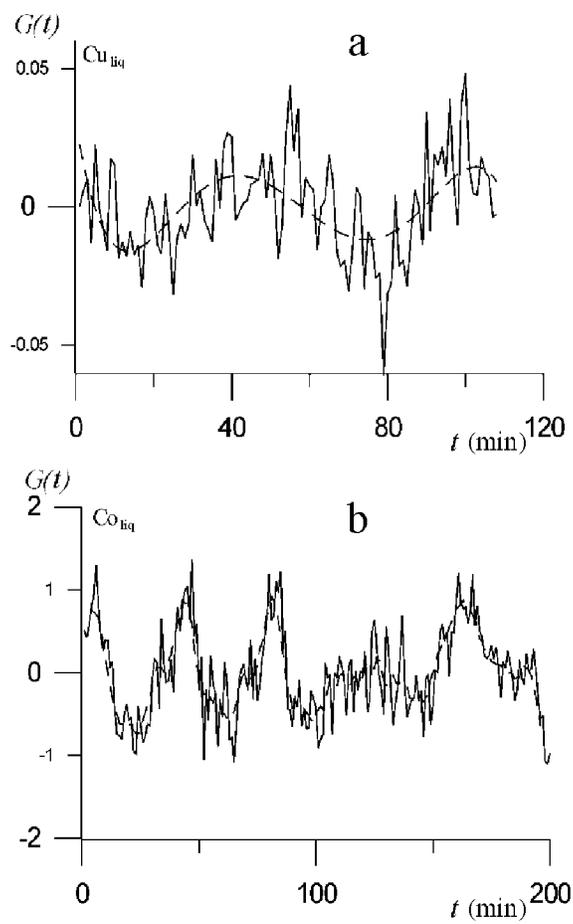}\caption{ Time dependences of autocorrelation function for the
liquid copper (a), and liquid cobalt (b) viscosity close to corresponding temperatures of step change of viscosity.}
\end{figure}

\begin{figure}[h]
   \centering
   \includegraphics[scale=1]{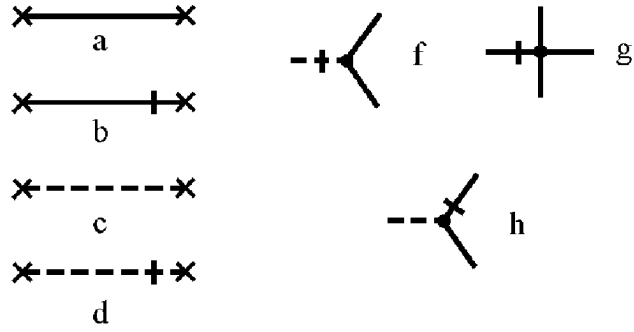}\caption{Graphs $a, b, c, d$  correspond to propagators $< \psi
\psi >$, $< \psi \psi '>$, $< v v >$, and $< v v'>$
 accordingly, graphs $f, g, h$  correspond to vertexes $v' \psi \partial (\partial ^2\psi )$, $\lambda g\psi '\partial
 ^2(\psi ^3)$, and $\psi 'v\partial \psi $  accordingly.}
\end{figure}

\begin{figure}[h]
   \centering
   \includegraphics[scale=1]{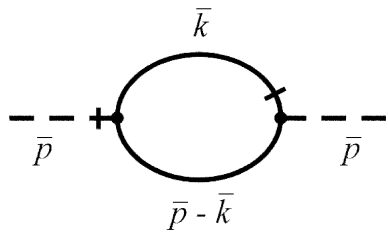}\caption{The graph corresponds to the response function $\Sigma
_{v'v}$ in the framework one-loop approximation.}
\end{figure}

\begin{figure}[h]
   \centering
   \includegraphics[scale=1]{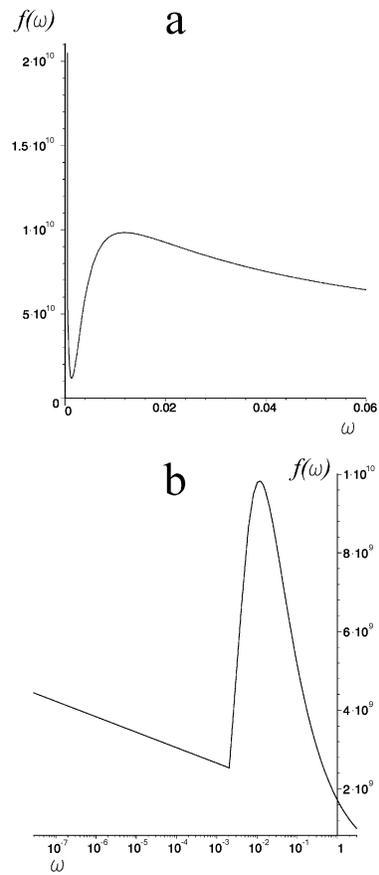}\caption{The spectrum of viscosity oscillations ($f(\omega )=\sqrt
{f_1(\omega )^2+f_2(\omega )^2}$) in linear--linear (a) and logarithmic--linear (b) coordinates.}
\end{figure}

\end{document}